\documentclass[%
 reprint,
superscriptaddress,
nofootinbib,
 amsmath,amssymb,
 aps,
 pra,
]{revtex4-1}
\usepackage{verbatim}
\usepackage{mathtools}
\usepackage{amsthm}
\usepackage{physics}
\usepackage{graphicx}
\usepackage{dcolumn}
\usepackage{bm}
\usepackage{xfrac}
\graphicspath{ {./images/} }

\begin{document}


\title{Two-orbital quantum discord in fermion systems}

\author{Javier Faba}

\affiliation{%
Center for Computational Simulation, Universidad Polit\'ecnica de Madrid, Campus Montegancedo, 28660 Boadilla del Monte, Madrid, Spain
}%

\affiliation{%
Departamento de F\'isica Te\'orica and CIAFF, Universidad Aut\'onoma de Madrid, E-28049 Madrid, Spain 
}%

\author{Vicente Mart\'\i n}

\affiliation{%
Center for Computational Simulation, Universidad Polit\'ecnica de Madrid, Campus Montegancedo, 28660 Boadilla del Monte, Madrid, Spain
}%

\author{Luis Robledo}
\affiliation{%
Departamento de F\'isica Te\'orica and CIAFF, Universidad Aut\'onoma de Madrid, E-28049 Madrid, Spain 
}%

\affiliation{%
Center for Computational Simulation, Universidad Polit\'ecnica de Madrid, Campus Montegancedo, 28660 Boadilla del Monte, Madrid, Spain
}%

\date{\today}

\begin{abstract}
A simple expression to compute the quantum discord between two orbitals in fermion systems is derived using the parity superselection rule. As the correlation between orbitals depends on the basis chosen, we discuss a special orbital basis, the natural one. We show that quantum correlations between natural orbital pairs disappear when the pairing tensor is zero, i.e. the particle number symmetry is preserved. The Hartree-Fock orbitals within a Slater determinant state, a Hartree-Fock-Bogoliubov quasiparticle orbitals in a quasiparticle vacuum, or the ground state of a Hamiltonian with particle symmetry and their corresponding natural orbitals are some relevant examples of natural basis and their corresponding states. Since  natural orbitals have that special property, we seek for the quantum discord in non-natural orbital basis. We analyze our findings in the context of the Lipkin-Meshkov-Glick and Agassi models.
\end{abstract}

\pacs{Valid PACS appear here}
\maketitle


\section{\label{sec:intro}Introduction}

Quantum correlations have been a central field of research since the inception of quantum mechanics  \cite{einstein_podolsky,schrodinger_1935}. They are a fundamental feature of the quantum theory and give rise to many interesting phenomena like those observed in the fields of quantum cryptography \cite{nielsen_chuang,quantum_cryptography,QuantumSystemsChannelsInformation,watrous_2018}, quantum teleportation \cite{teleportation1,teleportation2,teleportation3}, quantum phase transitions in many body systems \cite{sachdev2001quantum,quantum_phase_transitions_review,bond_charge_Hubbard}, etc.

They can be studied from different points of view. For instance, 
from a many body perspective, it is known that, if we solve a many body Hamiltonian through mean field techniques, such as the Hartree-Fock or Hartree-Fock-Bogoliubov method, the ground state will, in general, not preserve the symmetries of the Hamiltonian \cite{ring_schuck}. This spontaneous breaking of symmetries is interpreted as a way  to catch important correlations of the exact ground state while preserving the simple mean field picture. From a quantum information point of view, quantum correlations show up when we analyze the state of a given partition in a quantum system. For instance, if we have a pure state living in a Hilbert space with some tensor product structure, we can use the von Neumann entropy of one of the marginals in order to quantify the amount of entanglement between parties \cite{nielsen_chuang}.
However, when we try to join both perspectives, some subtleties arise. For example, when we try to quantify quantum correlations through partial traces, we usually need a Hilbert space with a tensor product structure. Nonetheless, in dealing with many body systems it is customary to deal with identical fermionic particles, and due to the antisymmetrization principle, the tensor structure is lost \cite{reasonable}. Moreover, as it is physically impossible to distinguish among identical particles, it is inconsistent to compute correlations between them through, for example, the von Neumann entropy of the marginals. While it's true that those subtleties are solved by defining a fermionic partial trace \cite{friis}, or quantifying the particle entanglement through the one body entropy \cite{yoshiko}, in general quantum information concepts are not directly applicable in fermionic many body systems. Some efforts with Helium-like systems have been done in \cite{helium_like_systems,helium_like_systems_2} and references therein. Moreover, concepts as the multipartite concurrence under the context of identical particles \cite{multipartite_concurrence}, the Lo Franco and Compagno’s approach \cite{LoFranco2016,entanglement_indistinguisahble} or the antisymmetric negativity \cite{experimental_entanglement_antisymmetrization} have been proposed, among others.

A very powerful measure of quantum correlations is the so-called quantum discord \cite{quantum_discord}. It quantifies all quantum correlations beyond entanglement and allows us to differentiate between classical and quantum correlations \cite{quantum_discord}. However, its calculation requires to perform arbitrary projective measurements in one of the subsystems and therefore it implicitly requires the existence of a tensor product structure. Moreover, it is in general hard to compute, both analytically and numerically, due to the variational process implicit in its definition \cite{discord_hard_analytically,discord_hard_numerically}.

In this work, we aim to study correlations in many body fermionic systems from a quantum information perspective through the analysis of the quantum discord. Specifically, we derive a very simple expression in order to compute the quantum discord between two fermionic orbitals in a arbitrary mixed state, and we apply it the characterization of some well known benchmarking models. The paper is structured as follows. In Sec. \ref{sec:QD} we briefly introduce the concept of quantum discord. We refer to \cite{review_quantum_discord,Adesso_2016} for a complete review of the quantum discord concept and quantum correlations. In Sec. \ref{sec:two_orbital} we derive an expression to compute the quantum discord between pair of orbitals in fermion systems. In \ref{sec:natural_orbitals} and \ref{sec:HF_HFB} we discuss some properties of quantum discord related to the orbital basis. In \ref{sec:results} we compute it in the context of the Agassi and Lipkin-Meshkov-Glick models. Finally, in Sec. \ref{sec:conclusions} and the appendix we summarize the results obtained and we discuss some connections with other results in the literature, respectively.

\section{\label{sec:QD}Quantum Discord}

Given the Hilbert space $\mathcal{H}$ of a quantum system, let us assume there exists a bi-partition 
 $\mathcal{H} = \mathcal{H^{(A)}} \otimes  \mathcal{H^{(B)}}$. The quantum discord, introduced by Ollivier and Zurek \cite{quantum_discord}, is a measure of the purely-quantum correlations beyond entanglement between both parts $A$ and $B$. It is defined as the discrepancy between two classically equivalent measures $I(A,B)$ and $J(A,B)$ of the mutual information
\begin{equation*}
\delta(A,B) = I(A,B) - J(A,B).
\end{equation*}
They are given by
\begin{equation}
\label{eq:I}
I(A,B) = S(\rho^{(A)})+S(\rho^{(B)})-S(\rho^{(A,B)})
\end{equation}
and
\begin{equation}
\label{eq:J}
J(A,B) = \max_{\{\Pi_{k}^{(B)}\}} S(\rho^{(A)})-S(\rho^{(A,B)}|\{\Pi_{k}^{(B)}\})
\end{equation}
While $I(A,B)$ is a measure of all kind of correlations, $J(A,B)$ quantifies the classical part. The measurement-based conditional entropy entering the definition of  $J(A,B)$ is defined as
\begin{equation*}
S(\rho^{(A,B)}|\{\Pi_{k}^{(B)}\}) = \sum_{k} p_k S(\rho_k^{(A,B)})
\end{equation*}
where $\rho_k^{(A,B)} = \frac{1}{p_k}\Pi_{k}^{(B)}\rho^{(A,B)}\Pi_{k}^{(B)}$ is the measured-projected state and $p_k = \tr (\Pi_{k}^{(B)}\rho^{(A,B)}\Pi_{k}^{(B)})$ is the associated probability. The measurement and the associated projector $\Pi_{k}^{(B)}$ are defined only in the sector $B$ of the bi-partition.

 Due to the variational process involved in Eq. (\ref{eq:J}) the quantum discord is hard to compute in general either  analytically \cite{discord_hard_analytically} or numerically \cite{discord_hard_numerically}. Some results exist for two qubit systems \cite{two_qubit, x-states}, qubit-qudit systems \cite{qubit_qudit}, and there is also some work related to the quantum discord in fermionic systems \cite{superconducting_systems, majorana_fermions, PhysRevA.100.052331, bond_charge_Hubbard}. The calculation is simplified (that is, there is no variational process involved) if there exist in the model considered some kind of selection rule that reduces drastically the variational space (i.e. the set of valid projective measurements in $B$). In this work, we derive a very simple expression for the two orbital quantum discord in a general fermionic system by using a number-parity selection rule, and we apply it under the context of the Agassi model. Although this type of approach is a novelty, similar equations were obtained in \cite{information_loss} considering the information loss due to a measurement of a single mode in a fermion system. 

\section{\label{sec:two_orbital}Two-Orbital fermionic system}

Consider a system formed by $\Omega$ orbitals occupied by fermions where number-parity symmetry \cite{ring_shuck} is preserved and can be considered as a selection rule (NPSR) \footnote{Wave functions mixing configurations with an even and odd number of fermions are not allowed.}. Since we are dealing with fermions, the single-orbital occupation may be $0$ (if there is no fermion in the orbital) or $1$ (if there is a single fermion in the orbital). We divide the system in three subsystems: $A$, $B$ and $C$. $A$($B$) corresponds to the $i$($j$)-th orbital, and C corresponds to the orthogonal complement of $AB$. Since all pure and mixed states must fulfill the NPSR, the density matrix corresponding to the $AB$ subsystem will have the following structure in the occupation basis:
\begin{equation}
\label{eq:rho_AB}
\rho^{(A,B)} = 
\begin{pmatrix}
\rho_{1} & 0 & 0 & \alpha \\
0 & \rho_{2} & \gamma & 0 \\
0 & \gamma^* & \rho_{3} & 0 \\
\alpha^* & 0 & 0 & \rho_{4} \\
\end{pmatrix}
\end{equation}
with $\sum_{i = 1}^4 \rho_{i} = 1$. Now we must find a complete set of projectors in the $B$ subspace. For this purpose, one would be tempted to follow the path as in \cite{two_qubit}, this is, performing $U(2)$ rotations on the two ``computational" local projectors. Nonetheless, as the NPSR must be fulfilled, a projective measurement that mix the occupied and non-occupied state of just one orbital would be unphysical. In fact, a self-adjoint operator must commute with the super-selection rule in order to be an observable. A measurement that do not respect the super-selection rule can not be related to any observable, so it would be unrealizable. Indeed, ignoring superselection rules, under the context of fermionic quantum information measures, could lead to a vast overestimation of the correlation or entanglement of the system \cite{ding_1,ding_2}. Then, the only possible projectors in the $j$-th orbital's occupation space are
\begin{equation}
\label{eq:projectors}
\begin{split}
\Pi_0 = a_ja^\dagger_j \\
\Pi_1 = a^\dagger_ja_j
\end{split}
\end{equation}
Since the set of possible projective measurements for part $B$ has just one pair of elements instead of infinite, no optimization process is involved in Eq. (\ref{eq:J}) and  quantum discord can be easily computed. The measured-projected states will be
\begin{equation*}
\begin{split}
\rho_0^{(A,B)} = \frac{1}{\rho_{1}+\rho_{3}}
\begin{pmatrix}
\rho_{1} & 0 & 0 & 0 \\
0 & 0 & 0 & 0 \\
0 & 0 & \rho_{3} & 0 \\
0 & 0 & 0 & 0 \\
\end{pmatrix} \\
\rho_1^{(A,B)} = \frac{1}{\rho_{2}+\rho_{4}}
\begin{pmatrix}
0 & 0 & 0 & 0 \\
0 & \rho_{2} & 0 & 0 \\
0 & 0 & 0 & 0 \\
0 & 0 & 0 & \rho_{4} \\
\end{pmatrix}
\end{split}
\end{equation*}
and, straightforwardly, the conditional entropy can be written as
\begin{equation*}
S(\rho^{(A,B)}|\{\Pi_{k}^{(B)}\}) = S\big(\mathcal{Z}(\rho^{(A,B)})\big)-S(\rho^{(B)})
\end{equation*}
where $\mathcal{Z}(\rho)$ is the de-phasing channel, i.e, the quantum channel that destroys the off-diagonal elements of $\rho$. Finally the quantum discord can be written as
\begin{equation}
\label{eq:quantum_discord}
\delta(A,B) = S\big(\mathcal{Z}(\rho^{(A,B)})\big) - S(\rho^{(A,B)})
\end{equation}
or, more explicitly, in terms of the two-orbital reduced matrix elements:
\begin{equation*}
\delta(A,B) = \sum_k \lambda_k\ln\lambda_k-\rho_{k}\ln \rho_{k}
\end{equation*}
with
\begin{equation}
\label{eq:eigenvalues_rho_AB}
\left\lbrace
\begin{aligned}
\lambda_0 &= \frac{\rho_{1}+\rho_{4}}{2}+\sqrt{\bigg(\frac{\rho_{1}-\rho_{4}}{2}\bigg)^2+\abs{\alpha}^2} \\
\lambda_1 &= \frac{\rho_{1}+\rho_{4}}{2}-\sqrt{\bigg(\frac{\rho_{1}-\rho_{4}}{2}\bigg)^2+\abs{\alpha}^2} \\
\lambda_2 &= \frac{\rho_{2}+\rho_{3}}{2}+\sqrt{\bigg(\frac{\rho_{2}-\rho_{3}}{2}\bigg)^2+\abs{\gamma}^2} \\
\lambda_3 &= \frac{\rho_{2}+\rho_{3}}{2}-\sqrt{\bigg(\frac{\rho_{2}-\rho_{3}}{2}\bigg)^2+\abs{\gamma}^2}
\end{aligned}
\right.
\end{equation}
The quantum discord between the $i$ and $j$ orbitals grows with the off-diagonal matrix elements of $\rho^{(A,B)}$ in Eq \ref{eq:rho_AB}, whose value reflects the amount of quantum coherence of the state. This fact can be easily seen if we write a pair of diagonal elements, i.e, $\rho_{1}$ and $\rho_{4}$ (the same discussion applies to $\rho_{2}$ and $\rho_{3}$) as $\rho_{1} = \frac{1}{4}+\epsilon$ and  $\rho_{4} = \frac{1}{4}-\epsilon$ with $\epsilon \in [0,\frac{1}{4}]$. This corresponds to the case in which $S\big(\mathcal{Z}(\rho^{(A,B)})\big)$ is maximum, and we perturb it via the parameter $\epsilon$. The eigenvalues are, in this case, $\lambda_0 = \frac{1}{4}+\sqrt{\epsilon^2+\abs{\alpha}^2}$ and $\lambda_1 = \frac{1}{4}-\sqrt{\epsilon^2+\abs{\alpha}^2}$. For all allowed values of $\epsilon$, $S(\rho^{(A,B)})\leq S\big(\mathcal{Z}(\rho^{(A,B)})\big)$ with equality when the off-diagonal elements are zero (in this case, $\alpha = 0$). If $\alpha$ increases, then the quantum discord increases too, revealing that the exclusively quantum correlations increase as the coherence grows, which is an expected and intuitive result. 

\section{\label{sec:natural_orbitals}Quantum discord and natural orbitals}

The two-orbital reduced density matrix $\rho^{(A,B)}$ can be written in terms of three well known quantities in many-body theory: the one-body density matrix, the two-body density matrix and the pairing tensor, defined as $\gamma_{i,j} = \langle a^\dagger_ja_i\rangle$,  $\gamma_{i,j,i,j} = \langle a^\dagger_ia^\dagger_ja_ja_i\rangle$ and $\kappa_{i,j} = \langle a_ja_i\rangle$, respectively. It has the same structure as Eq. (\ref{eq:rho_AB}), with:
\begin{equation}
\label{eq:rho_AB_with_densities}
\left\lbrace
\begin{aligned}
\rho_{1} &= 1-\gamma_{i,i}-\gamma_{j,j}+\gamma_{i,j,i,j} \\
\rho_{2} &= \gamma_{j,j}-\gamma_{i,j,i,j} \\
\rho_{3} &= \gamma_{i,i}-\gamma_{i,j,i,j} \\
\rho_{4} &= \gamma_{i,j,i,j} \\
\alpha &= \kappa_{j,i}^* \\
\gamma &= \gamma_{j,i}
\end{aligned}\right.
\end{equation}
Together with Eq. (\ref{eq:quantum_discord}), we see that the off-diagonal elements of the one-body density matrix and the pairing tensor are directly related with the quantum discord between $i$ and $j$ orbitals: if at least one of them is non-zero, there exist quantum correlations between $i$ and $j$. Inversely, it can be easily seen that there are two conditions for the quantum discord to be zero for all pairs of orbitals: $\gamma_{j,i} = 0$ and $\kappa_{j,i} = 0$. According with the results obtained in \cite{information_loss}, the first condition, $\gamma_{j,i} = 0$, is fulfilled for all $i,j$ if and only if the orbitals are the natural ones, i.e, those who diagonalize the one-body density matrix. Additionally, if the state commutes with the particle number operator, then $\kappa_{j,i} = 0$ for all $i,j$. Thus, the two conditions for the vanishing of quantum discord between all orbital pairs are:
\begin{enumerate}
    \item The orbital basis is the natural one
    \item The state commutes with the particle number operator
\end{enumerate} 

Additionally, it is known that the natural orbitals are the ones that minimize the overall entropy, defined as the sum of all the one-orbital entropies \cite{rosignolli}. Since this quantity is used to quantify the amount of total correlation in a state (the total entanglement if the state is pure) \cite{szalay}, then, if the number of particles is well defined and the state is pure, a non-zero overall entropy implies that all correlations between pairs of natural orbitals will be purely classical (if they exists) and the entanglement must be manifested between three orbitals or more.

\section{\label{sec:HF_HFB}General orbital basis}

It is important to remark that the quantum discord is measured between orbitals, and not between particles. For this reason, a change in the orbital basis may induce a change in the correlations between them. So, in order to study the quantum discord of a state, it is fundamental to specify properly the orbital basis. A natural orbital basis of a state which commutes with the particle number operator implies that those orbitals are constructed so that they can keep the intrinsic quantum correlation of the state without needing quantum correlation by pairs between them (this will be clearer in Sec. \ref{sec:lipkin}). Therefore the following question arises: which is the value of the quantum discord of a given state in a general orbital basis?

Suppose that we have a general orbital basis and the natural orbital basis (of the given state), related by the most general linear canonical transformation between creation/annihilation operators (Bogoliubov transformation \cite{ring_schuck}):
\begin{equation*}
\beta_k^\dagger = \sum_l U_{l,k}c^\dagger_l+V_{l,k}c_l
\end{equation*}
where $\{\beta_k^\dagger\}$ are the fermionic creation operators for the general basis, $\{c^\dagger_l\}$ are the fermionic creation operators for the natural basis. The following relations among the Bogoliubov amplitudes $U$ and $V$
\begin{equation}
\label{eq:U_V_relations}
\begin{split}
U^\dagger U+V^\dagger V &=U U^\dagger + V^* V^T = I \\
U^T V + V^T U &= UV^\dagger + V^* U^T = 0 \\
\end{split}
\end{equation}
hold. Then, the one-body matrix and the pairing tensor elements read \cite{ring_schuck}:
\begin{equation*}
\begin{split}
\gamma_{k,k'} &= \sum_lV^\dagger_{k,l}V_{l,k'}+(U^\dagger_{k,l}U_{l,k'}-V^\dagger_{k,l}V_{l,k'})p_l \\
\kappa_{k,k'} &= \sum_lV^\dagger_{k,l}U^*_{l,k'}+(U^\dagger_{k,l}V^*_{l,k'}-V^\dagger_{k,l}U^*_{l,k'})p_l \\
\end{split}
\end{equation*}
where $p_l = \langle c^\dagger_lc_l \rangle$. In this general case, the quantum discord will be nonzero and orbital dependent except for the case $p_l = \frac{1}{2}$ $\forall l$. This can be checked using the relations in (\ref{eq:U_V_relations}). Since $\sum_lp_l = N$ where $N$ is the number of particles, this case can only exist when the number of particles is exactly half the number of orbitals. Less restrictive is the case of a  transformation among particles, i.e, $V = 0$. Again, the quantum discord will be nonzero except for the case $p_l = \frac{N}{\Omega}$ $\forall l$ where $\Omega$ is the number of orbitals. In both cases, the quantum discord will be zero and orbital independent if the occupation of the system in the natural orbital basis is equally distributed. 

A common measure of entanglement among particles is the entanglement entropy of the one-body density matrix, defined as $S(\gamma) = -\gamma\log\gamma = -\sum_lp_l\log p_l$ \cite{yoshiko}. It is interesting to note that the orbital-independent zero discord case in general corresponds with the maximum $S(\gamma)$, which is indeed related to the overall entropy in the natural orbital basis since $S_{ov}^{nat} = -\sum_l p_l\ln p_l -\sum_l (1-p_l)\ln (1-p_l)$ \cite{rosignolli}, and it reaches its maximum value when $p_l = \frac{1}{2}$. In other words, for a pure system with half filling, i.e, $\Omega = 2N$, if the particles are equally distributed between all natural orbitals, the entanglement is maximum but there is no quantum correlations between orbital pairs.

\section{\label{sec:results}Results}

As an example of how the quantum discord between pairs of fermionic orbitals can be used to characterize the correlations in the system, we  apply the previous concepts to the HFB ground state of the Agassi model and the exact ground state of the Lipkin-Meshkov-Glick (LMG) model. Both models are composed by a two-level fermionic system, each having a $\Omega$-fold degeneracy. The difference between them lies in the interaction terms of their respective Hamiltonians.

The LMG model \cite{lipkin} has been widely used over the years as a benchmark in the characterization of different approximations to the many-body problem. The model is simple enough to be exactly solvable and at the same time is sufficiently rich to catch some non-trivial properties of  many-body systems, mainly, the quantum phase transition to a `deformed' state through a spontaneous symmetry breaking of the mean field approximation. As we will see, its Hamiltonian is composed by two terms: the non-interacting one, and the so-called monopole-monopole interaction, which mixes the high and low-lying orbitals of the same degeneracy. The model is very well known in the nuclear physics literature -- see Ref \cite{robledo92} for a study of the model adequate for the present purposes. Also, their entanglement properties have been widely studied \cite{vidal_1,lipkin_rossignoli}, as well as their correlation properties in a finite temperature context \cite{vidal_2} and the solution under the thermodynamical limit \cite{vidal_3,vidal_4}.

The Agassi model \cite{agassi}, is an extension of the LMG one where a separable pairing interaction has been added. The pairing interaction induces the creation and annihilation of particles by pairs with the same (and different) energies. When treated at the mean field level, the Agassi model contains a superfluid phase (treated using the Bardeen-Cooper-Schriefer theory of superconductivity) as well as a deformed one where the broken symmetry is parity (see \cite{robledo92} for a thorough discussion). The model can also be solved exactly using group theory techniques and it is also often used as a benchmark of different approximations in the context of nuclear physics.

\subsection{\label{sec:Agassi}The HFB ground state of the Agassi model as a benchmark}

The Agassi model \cite{agassi} is a two-level system, each of them with a degeneracy $\Omega$ (even). The system is filled with $N=\Omega$ fermions, and the Hamiltonian is given by
\begin{equation}
\label{eq:agassi_hamiltonian}
H = \epsilon J_0 -g\sum_{\sigma, \sigma^\prime}A_\sigma^\dagger A_{\sigma^\prime}
-\frac{1}{2}V[ (J_+)^2+(J_-)^2]
\end{equation}
with
\begin{equation}
\label{eq:SO5_generators}
\begin{split}
J_0 &= \frac{1}{2}\sum_{\sigma, m}\sigma c^\dagger_{\sigma, m}c_{\sigma, m} \\
J_+ &= (J_-)^\dagger = \sum_m c^\dagger_{1,m}c_{-1,m} \\
A_\sigma &= \sum_{m>0} c_{\sigma,-m}c_{\sigma,m}
\end{split}
\end{equation}
where $\sigma = \pm 1$ labels the upper/lower level, $m = \pm 1, \pm 2, ..., \pm \frac{\Omega}{2}$ labels the states within a level, and $c^\dagger_{\sigma, m}$ is the fermionic creation operator of the single particle state labelled by $(\sigma,m)$\footnote{Those states are also called `Hamiltonian orbitals' throughout this work.}. This model is exactly solvable using group theory methods \cite{agassi_solution}, and the HFB ground state solution can be easily obtained. For this reason, we are going to analyze the quantum correlation properties of the HFB ground state as a benchmark of the proposed measure of quantum discord between pairs of fermionic orbitals (Eq. (\ref{eq:quantum_discord})).

Following reference \cite{agassi_solution}, the one body density matrix and the pairing tensor of the HFB ground state can be written as
\begin{equation}
\label{eq:variables_agassi}
\begin{split}
\gamma_{\sigma m,\sigma^\prime m^\prime} &= \gamma_{\sigma,\sigma^\prime}\delta_{m,m^\prime} \\
\kappa_{\sigma m,\sigma^\prime m^\prime} &= \text{sgn}(m)\frac{1}{2}\sin \alpha \delta_{\sigma,\sigma^\prime} \delta_{m,-m^\prime}
\end{split}
\end{equation}
with 
\begin{equation}
\label{eq:variables_agassi_2}
\begin{split}
\gamma_{\sigma,\sigma} &= \frac{1}{2}(1-\sigma \cos\phi\cos\alpha) \\
\gamma_{\sigma,-\sigma} &= -\frac{1}{2}\sin\phi\cos\alpha
\end{split}
\end{equation}
The values of $\phi$ and $\alpha$ depend on the parameters of the Hamiltonian, that is:
\begin{equation*}
     \begin{cases}
       \phi = \alpha = 0 &\quad \text{if } \chi, \Sigma_0<1\\
       \cos\phi = \frac{1}{\chi}, \alpha = 0 &\quad \text{if } \chi>\Sigma_0\\
       \phi = 0, \cos\alpha = \frac{1}{\Sigma_0} &\quad \text{if } \chi<\Sigma_0\\
     \end{cases}
\end{equation*}
with $\chi = \frac{(\Omega-1)V}{\epsilon}$, $\Sigma = \frac{(\Omega-1)g}{\epsilon}$ and $\Sigma_0 = \Sigma +\frac{V}{\epsilon}$. As can be seen, there are three differentiated regions in the parameters space: the HF spherical phase, the HF deformed phase and the BCS phase. The first one corresponds to the conditions $\chi, \Sigma_0<1$, and the HFB ground state is the non interacting exact ground state, i.e, all the lower levels occupied. The second one corresponds to the conditions $\chi>\Sigma_0$ and $\chi>1$. In this case the HFB ground state breaks the parity symmetry\footnote{In the context of the Agassi model, particles in the upper (lower) level are assumed to have positive (negative) parity.} (that's why it's called `deformed'). The last region corresponds to $\chi<\Sigma_0$ and $\Sigma_0>1$. It preserves the parity symmetry but it breaks the particle number symmetry and represents a superfluid system described by the BCS approximation. Since in all regions the ground state is defined as a quasi-particle vacuum, the two body density is separable and the diagonal elements can be written as
\begin{equation*}
\gamma_{i,j,i,j} = \gamma_{i,i}\gamma_{j,j} + \Delta_{i,j}
\end{equation*}
with $\Delta_{i,j} = \abs{\kappa_{i,j}}^2-\abs{\gamma_{i,j}}^2$ and, using Eq. (\ref{eq:rho_AB_with_densities}) and (\ref{eq:rho_AB}), we can write the two orbital reduced density matrix as
\begin{equation*}
\left\lbrace
\begin{aligned}
\rho_{1} &= (1-\gamma_{i,i})(1-\gamma_{j,j}) + \Delta_{i,j} \\
\rho_{2} &= (1-\gamma_{i,i})\gamma_{j,j} - \Delta_{i,j}  \\
\rho_{3} &= \gamma_{i,i}(1-\gamma_{j,j}) - \Delta_{i,j} \\
\rho_{4} &= \gamma_{i,i}\gamma_{j,j} + \Delta_{i,j}  \\
\alpha &= \kappa_{j,i}^* \\
\gamma &= \gamma_{j,i} 
\end{aligned} \right .
\end{equation*}

\begin{figure}[ht]
\includegraphics[width=0.5\textwidth]{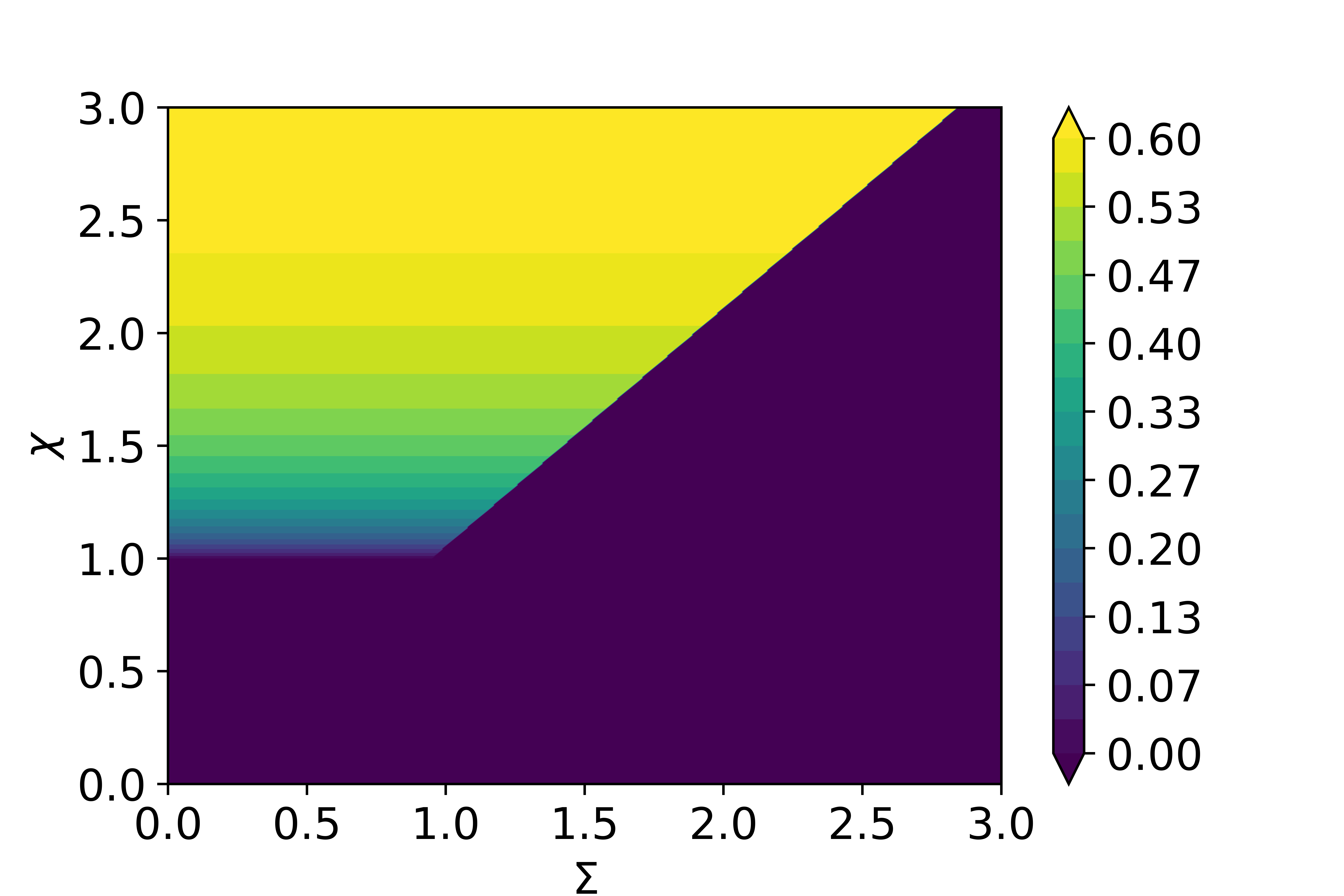}
\caption{Quantum discord between an orbital pair with $m=m^\prime$ and $\sigma = -\sigma^\prime$ as a function of the two Hamiltonian parameters $\chi$ and $\Sigma$. The quantum correlations in this case are zero for the spherical HF and BCS regions, while is non-zero only in the deformed HF region. $\Omega = 20$.}
\label{fig:up_down_quantum_discord}
\end{figure}

\begin{figure}[ht]
\includegraphics[width=0.5\textwidth]{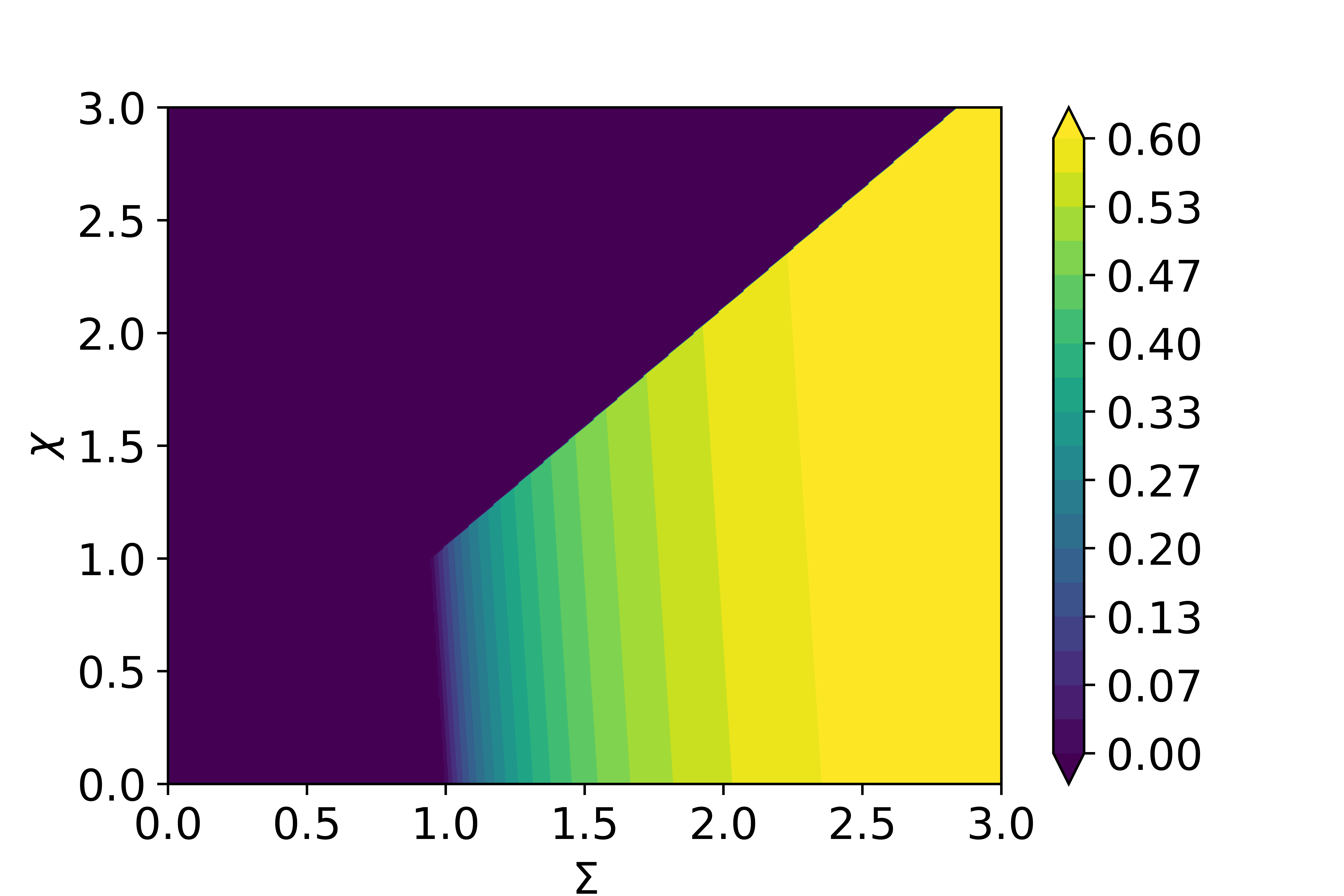}
\caption{Quantum discord between an orbital pair with $m=-m^\prime$ and $\sigma = \sigma^\prime$ as a function of the two Hamiltonian parameters $\chi$ and $\Sigma$. The quantum correlations in this case are zero for the spherical and deformed HF regions, while is non-zero only in the BCS region. $\Omega = 20$.}
\label{fig:pairing_quantum_discord}
\end{figure}

With those results and using Eq. (\ref{eq:quantum_discord}) we can easily compute the quantum discord between a pair of orbitals in the HFB ground state solution, which is
\begin{equation}
\label{eq:quantum_discord_agassi}
    \begin{cases}
        \delta(m,\sigma ; m,-\sigma) = h(\chi) &\quad \text{in the deformed HF region} \\
        \delta(m,\sigma ; -m,\sigma) = h(\Sigma_0) &\quad \text{in the BCS region} \\
        \delta(m,\sigma ; m^\prime, \sigma^\prime) = 0 &\quad \text{otherwise} \\
    \end{cases}
\end{equation}
with $h(x) = -\frac{1}{2}(1-\frac{1}{x})\ln \frac{1}{2}(1-\frac{1}{x})-\frac{1}{2}(1+\frac{1}{x})\ln \frac{1}{2}(1+\frac{1}{x})$. This solution is shown in Figs. \ref{fig:up_down_quantum_discord} and \ref{fig:pairing_quantum_discord}.

The structure of the quantum discord is the same as the phase diagram \cite{agassi_solution}. In the deformed HF phase, there are quantum correlations only between orbitals with the same $m$ and opposite $\sigma$ due to the monopole-monopole interaction. In the same way, there are quantum correlations in the BCS phase only between orbitals with the same $\sigma$ and opposite $m$ due to the pairing interaction in the Hamiltonian. In fact, if we compute the mutual information, defined in Eq. (\ref{eq:I}), we obtain $I(A,B) = 2\delta(A,B)$. This is expected when the state $\rho^{(A,B)}$ is pure \cite{two_qubit} and, indeed, it is the case within the HFB solution\footnote{It can be checked that the eigenvalues of the two-body density matrix, i.e, $\lambda_i$ in Eq. (\ref{eq:eigenvalues_rho_AB}) acquire the values $\lambda_i = 0,1$.}. Specifically, the two orbital reduced state between orbitals with the same $m$ and opposite $\sigma$ is pure in the deformed HF region and mixed in the BCS one (the inverse happens with same $\sigma$ and opposite $m$ orbitals). This result serves as a benchmark of the results obtained in Sec. \ref{sec:two_orbital}. Moreover, we note that the quantum discord between a transition from a spherical HF state to a deformed HF or BCS state is continuous, while the quantum discord between a transition from a deformed HF state to a BCS one is discontinuous. Since the quantities $\rho = -\frac{1}{2}\sin\phi\cos\alpha$ and $\kappa = \frac{1}{2}\sin\alpha$ from Eqs. (\ref{eq:variables_agassi}) and (\ref{eq:variables_agassi_2}) can be considered as order parameters of the model \cite{agassi_solution}, the quantum discord shows the behaviour of a combined order parameter.

It is interesting to note that there is no quantum discord between the Hamiltonian orbital pairs considered when the state is the exact ground state, since the corresponding one-body matrix elements are zero and the particle number is well defined (see Appendix \ref{sec:exact_agassi} for details). As we will discuss in next section, a low quantum discord implies a better adaptation of the orbitals in order to describe the state. This indicates that the Hamiltonian orbitals are suited to describe the exact ground state better than the HFB ground state, as expected.

\subsection{\label{sec:lipkin}The exact ground state of the LMG model}

Finally, we analyze the quantum discord between orbital pairs within the exact ground state of the LMG model. The LMG Hamiltonian is the same as Eq. (\ref{eq:agassi_hamiltonian}), with $g = 0$, i.e, there are only monopole-monopole interaction. For this reason, we only consider the quantum discord between orbitals with same $m$ and opposite $\sigma$.

Since the Hamiltonian commutes with the particle number and parity operators, the Hamiltonian orbitals, represented by the creation/annihilation operators ($c^\dagger_{\sigma, m}$ and $c_{\sigma, m}$ respectively) in Eq. (\ref{eq:agassi_hamiltonian}) are the natural ones ($\gamma_{i,j} = 0$ for $i\neq j$), and the pairing tensor is zero. Thus, as explained in Sec. \ref{sec:natural_orbitals}, the quantum discord is zero for all pairs.

\begin{figure}[ht]
\includegraphics[width=0.5\textwidth]{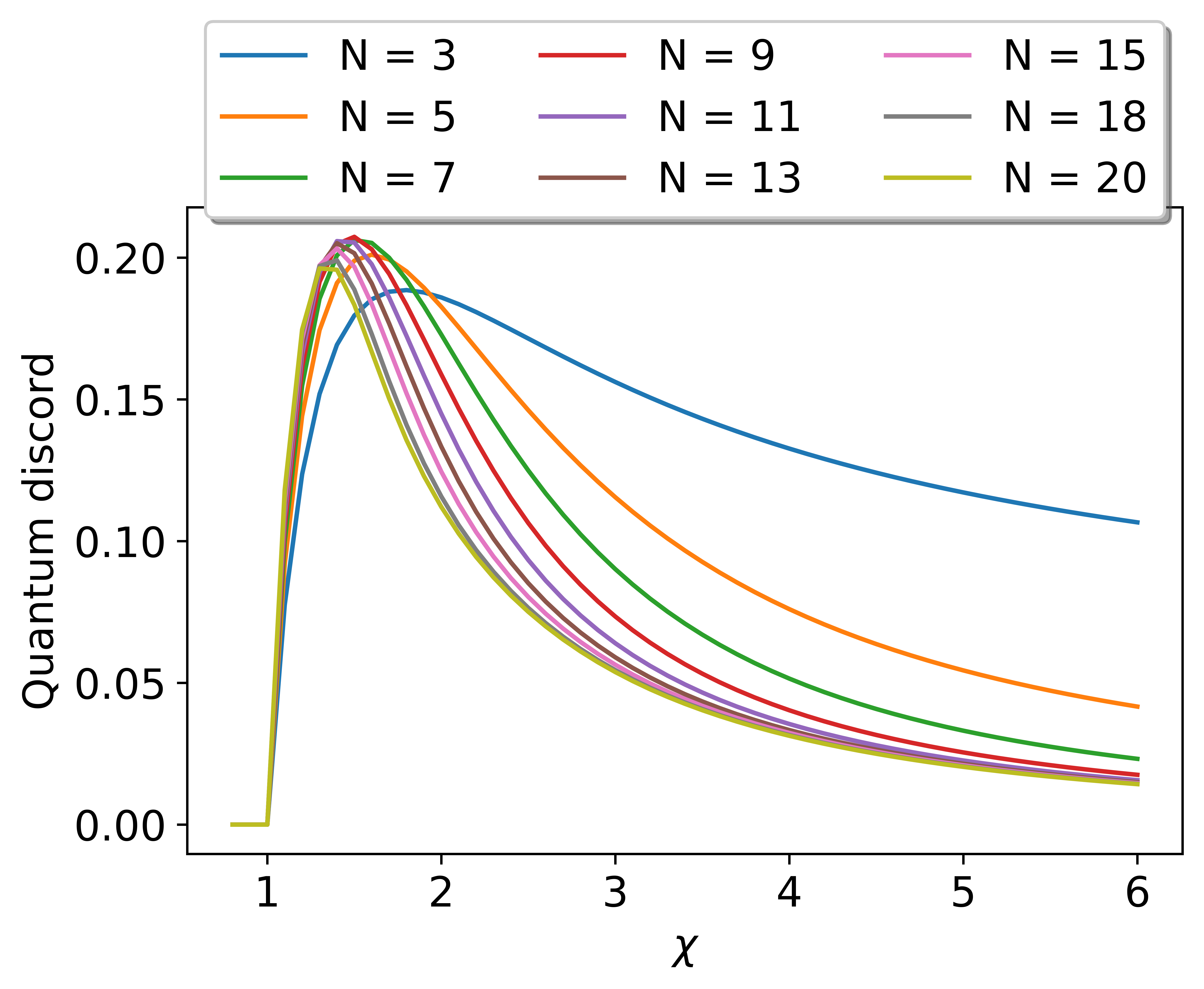}
\caption{Quantum discord between an up-down orbital pair as a function of the Hamiltonian parameter $\chi$ with $N$ from $3$ to $20$.}
\label{fig:QD_vs_kappa_Nmax_20}
\end{figure}

But this is not true if we change the orbital basis. In general, a low quantum discord implies a better adaptation of the orbitals in order to describe the exact ground state, while a high quantum discord reflects the contrary case. If we compute the quantum discord between an up-down pair of HF orbitals, we obtain the result shown in Fig. \ref{fig:QD_vs_kappa_Nmax_20}. It is interesting to analyze the behaviour of the quantum discord of the exact ground state between those orbitals since they are defined in order to catch the maximum correlations as possible within a mean field scenario.

For $\chi<1$ the quantum discord is zero, since the HF orbitals in this region coincide with the natural ones. For $\chi>1$ there are two different regimes. First, as $\chi$ is big enough and it grows, the quantum discord decreases. This decrease is more drastic if the particle number is bigger, in coherence with the mean field description, in which more accuracy is obtained when the number of particles is big enough.
The other regime is manifested when $\chi>1$ acquires intermediate values, this is, near the quantum phase transition point ($\chi=1$) and far from the asymptotic limit. In this region, the quantum discord grows fast until reaching the maximum. Then, it decreases exponentially until the asymptotic regime. This intermediate region is where the Hartree-Fock approximation becomes less accurate, and this is reflected as a high quantum discord between the HF orbitals: since the orbitals are less optimum in order to encode the exact ground state, more quantum correlation is needed between them for that task. In this intermediate region it is necessary to consider linear combinations of mean field Slater determinants to catch the physics of the exact ground state \cite{robledo92}.

\begin{figure}[ht]
\includegraphics[width=0.5\textwidth]{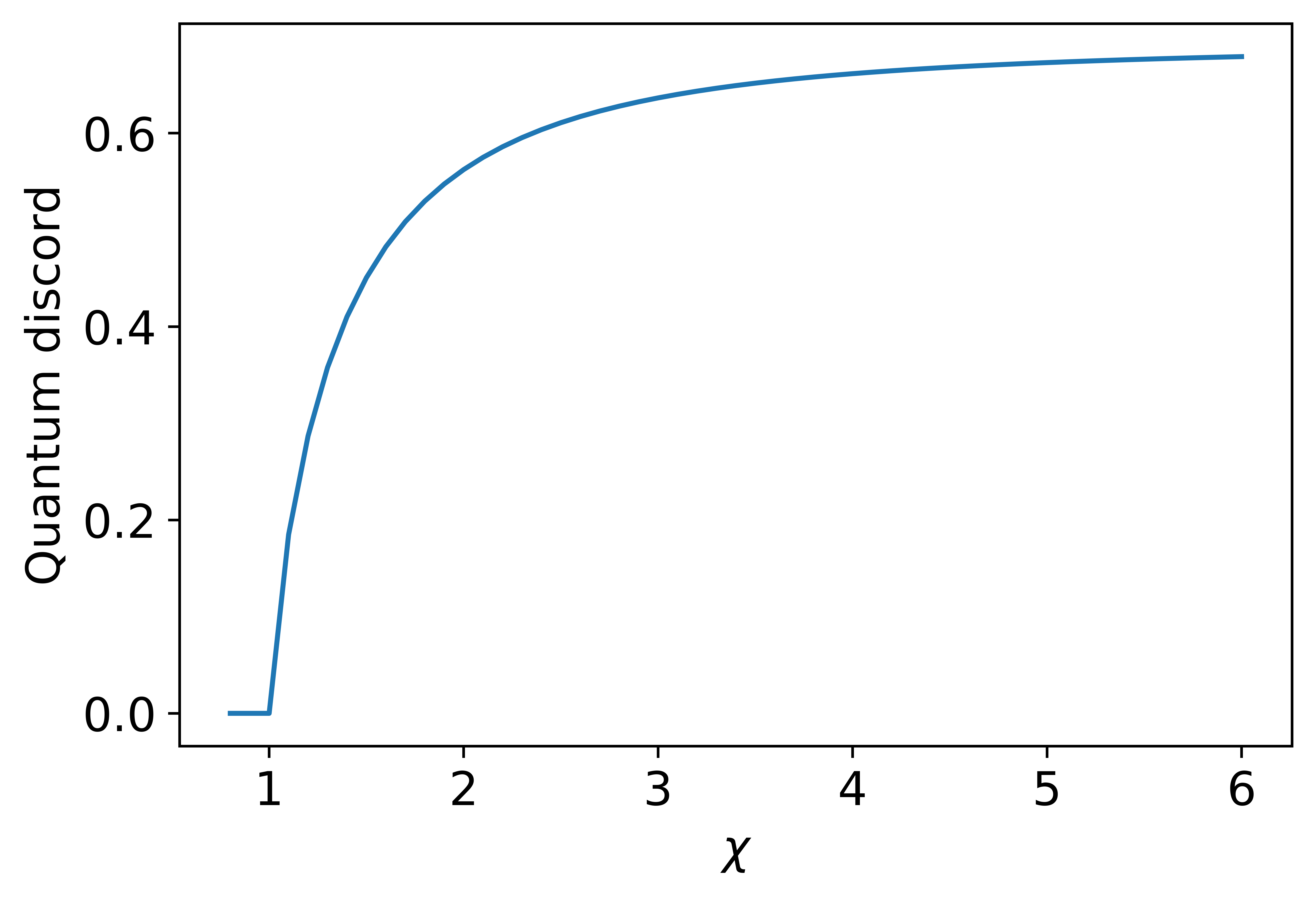}
\caption{Quantum discord between an up-down Hamiltonian orbital pair for the HF ground state as a function of the Hamiltonian parameter $\chi$.}
\label{fig:QD_vs_kappa_HF_GS}
\end{figure}

Until now, we have discussed the quantum discord between an up-down HF orbital pair for the exact ground state of the LMG model. We argued that, since the Hamiltonian orbitals are the natural ones, the same quantity between those is zero. The same argument apply for the HF orbitals in a HF ground state. However, we can ask ourselves what is the quantum discord between an up-down Hamiltonian orbital pair of the HF ground state (which is the `inverse' case with respect to the results in Fig. (\ref{fig:QD_vs_kappa_Nmax_20})). Since the LMG model is a particular case of the Agassi model, we find that this quantity is given by Eq. (\ref{eq:quantum_discord_agassi}) when $\Sigma = 0$ (Fig. (\ref{fig:QD_vs_kappa_HF_GS})).

Unlike Fig. (\ref{fig:QD_vs_kappa_Nmax_20}), now the quantum discord approaches to the value $\ln 2$ when $\chi \rightarrow \infty$ and doesn't depend on the particle number. This different behaviour is coherent with the fact that the HF ground state is rather different than the exact one: although the HF orbitals are better adapted when $\chi$ is high, as discussed previously, the structure of the HF ground state remains a Slater determinant, which in general is far from being exact. Thus, the Hamiltonian orbitals require higher quantum correlations in order to describe this state when the interaction is large. So, within the context of the models considered, a low quantum discord between pairs of orbitals can be related to an optimal orbital adaptation when describing a given state.

\section{\label{sec:conclusions}Conclusions}

The quantum discord is a measure of  quantum correlations in a given state. It is defined as the minimum difference between two classically equivalent but quantumly different versions of the mutual information. This definition is based in the fact that, given a bi-partition $A|B$ of a system, a measurement on $A$ may break the quantum correlations between $A$ and $B$. In this manner, a projective measurement may be performed in one of the subsystems. Nonetheless, a fermion system must satisfy the parity superselection rule, so not all the projective measurements are physical. 

In this work, we use this property and we propose a simple expression (see Eq (\ref{eq:quantum_discord})) in order to compute the quantum discord between two orbitals in a general fermionic pure or mixed state. This expression does not require an optimization procedure and is directly related to two central many-body quantities: the one body density matrix and the pairing tensor. Thus, we have shown that the natural orbital basis, which is defined as the one that diagonalize the one body density matrix, reduces the quantum discord between any pair of orbitals to zero when the state commutes with the particle number operator. Moreover, when the system's orbitals are half filled, there is no quantum correlations between pairs for any arbitrary orbital basis. Finally, we compute and discuss the quantum discord between pairs of orbitals in the HFB ground state of the Agassi model, and in the exact ground state of the LMG model. Our results may be useful in order to analyze quantum correlations in more complicated and realistic many body fermionic systems.

\acknowledgments{}
The authors want to thank the Madrid regional government, Comunidad 
Aut\'onoma de Madrid, for the project Quantum Information Technologies: 
QUITEMAD-CM P2018/TCS-4342.
The  work of LMR was supported by Spanish Ministry 
of Economy and Competitiveness (MINECO) Grants No. 
PGC2018-094583-B-I00.

\appendix
\section{Relation between two orbital fermionic quantum discord, variational diagonalization and measurement induced disturbance}
In this appendix we briefly comment some connections between the proposed measure of quantum discord and the literature.

Following Eq. (\ref{eq:quantum_discord}), and taking into account that $\delta(A,B) = 0 \Leftrightarrow \mathcal{Z}(\rho^{(A,B)}) = \rho^{(A,B)}$, the quantum discord can be used as a cost function for a variational quantum state diagonalization algorithm \cite{variational_diagonalization}. Indeed, following the inverse argument, the difference of purities between $\mathcal{Z}(\rho^{(A,B)})$ and $\rho^{(A,B)}$, which is used as a cost function in \cite{variational_diagonalization}, could be interpreted as a measure of the quantum correlations, at least, in the case of a two fermionic orbital state.

On the other hand, Shunlong Luo proposed in \cite{measurement_induced_disturbance} an alternative way to characterize the quantum correlations of a state. He defined classical states as the ones that fulfill the condition\footnote{It is assumed here that $\rho \in \mathcal{H}^{(A)}\otimes\mathcal{H}^{(B)}$} $\Pi(\rho) = \rho$ with $\Pi(\rho) = \sum_{k,l} \Pi_{k}^{(A)}\otimes\Pi_{l}^{(B)}\rho\Pi_{k}^{(A)}\otimes\Pi_{l}^{(B)}$ and $\Pi_{k}^{(A)}$, $\Pi_{k}^{(B)}$ general projective measurements into the $A$ and $B$ systems, respectively. As explained in Sec. \ref{sec:two_orbital}, the only possible projectors are those of Eq. (\ref{eq:projectors}). Thus, we have
\begin{equation*}
\Pi(\rho^{(A,B)}) = \mathcal{Z}\big(\rho^{(A,B)}\big)
\end{equation*}
and therefore, Eq.  (\ref{eq:quantum_discord}) could be interpreted as a distance between the dephased density matrix and the original one, and therefore, the quantum discord and the measurement-induced disturbance coincide.

\section{Multipartite generalization of quantum discord}

Until now, we have only taken into account quantum correlations among pairs of orbitals. In this appendix, we discuss the quantum discord beyond the bipartite case and derive an expression for a measure of the total quantum correlation of a state which, indeed, matches with the proposed generalization of the multipartite quantum discord in \cite{multipartite_quantum_discord}.

As explained in Sec \ref{sec:natural_orbitals}, the overall entropy is a measure of the total entanglement in a pure state (the total correlation if the state is mixed) \cite{szalay}. With the definition in Eq. (\ref{eq:I}), we can write the overall entropy as
\begin{equation*}
\begin{split}
S_{ov} &= I(\Omega-1;\Omega)+I(\Omega-2;\Omega-1,\Omega) \\
&+I(\Omega-3;\Omega-2,\Omega-1,\Omega)+...+S(\rho)
\end{split}
\end{equation*}
where $I(i;j,k,...,l)$ is the mutual information (Eq. (\ref{eq:I})) with $A$ as the $i$-th orbital and $B$ as the system composed by the $j,k,...,l$-th orbitals. $S(\rho)$ is the von Neumann entropy of the system's density matrix. Naturally, if the system is pure, $S(\rho) = 0$. Since the mutual information quantifies the total correlation, both classical and quantum, between parties, and the overall entropy measures the total correlation encoded in a state \cite{szalay}, then we propose the following quantity
\begin{equation*}
\begin{split}
S_{ov}' &= J(\Omega-1;\Omega)+J(\Omega-2;\Omega-1,\Omega) \\
&+J(\Omega-3;\Omega-2,\Omega-1,\Omega)+...+S(\rho)
\end{split}
\end{equation*}
as a measure of the total classical correlation encoded in a state. Then, the total quantum correlation, i.e, the multipartite generalization of the quantum discord of a state will be the difference between $S_{ov}$ and $S_{ov}'$. Since 
\begin{equation*}
    J(i;j,k,...,l)=\max_{\{ \Pi^{(j,k,...,l)}_\alpha\}} \bigg(S(i)-S(i,j,k,...,l|\{ \Pi^{(j,k,...,l)}_\alpha\})\bigg)
\end{equation*}
where $\Pi^{(j,k,...,l)}_\alpha$ are the $\alpha$-th projector living in the space formed by the $j,k,..,l$ orbitals, then we have
\begin{equation*}
\begin{split}
\delta(i,...,l) &= S_{ov}-S_{ov}' \\
&=\min_{\{\Pi_\alpha\}} S(m,l|\{\Pi_\alpha^{(l)}\})+S(k,m,l|\{\Pi_\alpha^{(m,l)}\}) \\
&+...-S(i,...,k,m,l|l)
\end{split}
\end{equation*}
as an expression for the multipartite quantum discord. This proposal coincides with the one in Ref \cite{multipartite_quantum_discord}. This alternative derivation justifies the validity of the result and, since it is related to the overall entropy, may be interesting in future work to study its relationship with the orbital basis used, as well as apply it in the study of several models. Of special interest would be to study the connection with the  natural basis, which is the one that minimizes $S_{ov}$ \cite{rosignolli}.

\section{Exact ground state quantum discord in the Agassi model}
\label{sec:exact_agassi}

The exact ground state of the Agassi model can be obtained easily using group theory arguments. Following Refs. \cite{agassi, R5}, all the operators in Eq. (\ref{eq:SO5_generators}) are part of the $SO(5)$ generators. In this way, the exact ground state and energy can be obtained diagonalizing the Hamiltonian (\ref{eq:agassi_hamiltonian}) in terms of the basis within a given irrep of $SO(5)$. Since $SU(2)\times SU(2) \subset SO(5)$, the elements of this basis can be labelled as $\{\ket{(J_m,\Lambda_m);J,M_J,\Lambda,M_\Lambda}\}$, where $(J_m,\Lambda_m)$ labels the irrep (it represents the maximum values of the angular momentums), and the pairs $(J,M_J)$ and $(\Lambda,M_\Lambda)$ behave as two independent angular momentum quantum numbers. We are interested in the irrep given by $J_m = \Lambda_m = \frac{\Omega}{4}$, since this one contains the half-filled non interacting ground state. The angular momentum quantum numbers are related to the number of particles and seniority\footnote{Here we call seniority to the number of unpaired states. A filled state labelled by $(\sigma,m)$ is unpaired when the state $(\sigma,-m)$ is unfilled. Otherwise, the state is paired.} of the upper and lower levels by
\begin{equation*}
\begin{split}
 N_- &= 2M_J+\frac{\Omega}{2} \qquad V_- = \frac{\Omega}{2}-2J \\
 N_+ &= 2M_\Lambda+\frac{\Omega}{2} \qquad V_+ = \frac{\Omega}{2}-2\Lambda
\end{split}
\end{equation*}
where $N_+$, $N_-$ denotes the number of particles in the upper/lower levels respectively, and $V_+$, $V_-$ denotes the seniority of the upper/lower levels respectively. For more details, we refer the reader to references \cite{agassi, R5}.

With this, in order to compute the quantum discord between the $\{(\sigma,m),(-\sigma,m)\}$ and $\{(\sigma,m),(\sigma,-m)\}$ levels, we must compute the one-body density matrix elements $\bra{GS}c^\dagger_{\sigma,m}c_{-\sigma,m}\ket{GS}$ and $\bra{GS}c^\dagger_{\sigma,m}c_{\sigma,-m}\ket{GS}$ where $\ket{GS}$ denotes the exact ground state of the Hamiltonian (\ref{eq:agassi_hamiltonian}). However, it can be seen that $\bra{GS}c^\dagger_{\sigma,m}c_{-\sigma,m}\ket{GS} = \bra{GS}c^\dagger_{\sigma,m}c_{\sigma,-m}\ket{GS} = 0$. For the $\{(\sigma,m),(-\sigma,m)\}$ levels, the reason is simple. Using the definitions in (\ref{eq:SO5_generators}) we can write
\begin{equation*}
    \bra{GS}c^\dagger_{\sigma,m}c_{-\sigma,m}\ket{GS} = \frac{\bra{GS}J_\sigma\ket{GS}}{\Omega}
\end{equation*}
since the value of $\bra{GS}c^\dagger_{\sigma,m}c_{-\sigma,m}\ket{GS}$ must be the same for all $m$. However, only the matrix elements of the Hamiltonian in (\ref{eq:agassi_hamiltonian}) which connect states that differ in their quantum numbers by zero or $\pm 1$ are nonzero. For this reason, the ground state can only be constructed as a linear superposition of integer or half-integer states. Since the operators $J_{\pm}$ only have nonzero elements between states that differ by $\pm \frac{1}{2}$ in their quantum numbers \cite{agassi}, then $\bra{GS}c^\dagger_{\sigma,m}c_{-\sigma,m}\ket{GS} = 0$.

Finally, we will justify $\bra{GS}c^\dagger_{\sigma,m}c_{\sigma,-m}\ket{GS} = 0$. We expand the ground state in terms of the $SO(5)$ basis\footnote{We have omitted the irrep label for simplicity.}:
\begin{equation*}
    \ket{GS} = \sum C_{J,M_J,\Lambda,M_\Lambda}\ket{J,M_J,\Lambda,M_\Lambda}
\end{equation*}
The one-body matrix element between the $\{(\sigma,m),(\sigma,-m)\}$ levels can be written as
\begin{equation*}
\begin{split}
    \bra{GS}c^\dagger_{\sigma,m}c_{\sigma,-m}\ket{GS} &= 
    \sum C_{J',M_J',\Lambda ',M_\Lambda '}^* C_{J,M_J,\Lambda,M_\Lambda} \\
    &\bra{J',M_J',\Lambda ',M_\Lambda '}c^\dagger_{\sigma,m}c_{\sigma,-m}\ket{J,M_J,\Lambda,M_\Lambda} 
\end{split}
\end{equation*}
Since the operator $c^\dagger_{\sigma,m}c_{\sigma,-m}$ doesn't change $N_\pm, V_\pm$, then
\begin{equation*}
    \begin{split}
    &\bra{J',M_J',\Lambda',M_\Lambda'}c^\dagger_{\sigma,m}c_{\sigma,-m}\ket{J,M_J,\Lambda,M_\Lambda} = \\
    &\bra{J,M_J,\Lambda,M_\Lambda}c^\dagger_{\sigma,m}c_{\sigma,-m}\ket{J,M_J,\Lambda,M_\Lambda}\delta_{J',J}\delta_{M_J',M_J}\delta_{\Lambda',\Lambda}\delta_{M_\Lambda',M_\Lambda}
    \end{split}
\end{equation*}
Now, we expand the $\ket{J,M_J,\Lambda,M_\Lambda}$ state in terms of the occupational basis, we fix the label $m$, and we analyze all the even\footnote{Since all the operators of the algebra of $SO(5)$ create/annihilate particles by pairs, the states with odd occupation don't exist.} occupational states for the reduced system formed by the four states $\{(\sigma,m),(-\sigma,m),(\sigma,-m),(-\sigma,-m)\}$:

1) Zero particles state: $c^\dagger_{\sigma,m}c_{\sigma,-m}\ket{}=0$

2) Two particle states: The particles can't occupy the levels $\{(\sigma,m),(-\sigma,m)\}$ since the algebra operators \cite{agassi} can only create/annihilate particles within the pairs $\{(\sigma,m),(\sigma,-m)\}$ and  $\{(\sigma,m),(-\sigma,-m)\}$. For the other possible combinations, $\langle c^\dagger_{\sigma,m}c_{\sigma,-m} \rangle = 0$.

3) Four particle state: $c^\dagger_{\sigma,m}c_{\sigma,-m}\ket{}=0$

With this, it can be seen that $\bra{J,M_J,\Lambda,M_\Lambda}c^\dagger_{\sigma,m}c_{\sigma,-m}\ket{J,M_J,\Lambda,M_\Lambda}= 0$ and therefore $\bra{GS}c^\dagger_{\sigma,m}c_{\sigma,-m}\ket{GS} = 0$.

\bibliography{bibliography}

\end{document}